\documentclass[twocolumn,showpacs,preprintnumbers,amsmath,amssymb]{revtex4-1}

\usepackage{graphicx}
\usepackage{dcolumn}
\usepackage{bm}
\usepackage{hyperref}

\begin{document}


\title{Transition from amplitude to oscillation death under mean-field diffusive coupling} 



\author{Tanmoy Banerjee}
\email{tbanerjee@phys.buruniv.ac.in}
\affiliation{Department of Physics, University of Burdwan, Burdwan 713 104, West Bengal, India.}
\author{Debarati Ghosh}
\affiliation{Department of Physics, University of Burdwan, Burdwan 713 104, West Bengal, India.}


\received{:to be included by reviewer}
\date{\today}

\begin{abstract}
We study the transition from amplitude death (AD) to oscillation death (OD) state in limit-cycle oscillators coupled through mean-field diffusion. We show that this coupling scheme can induce an important transition from AD to OD even in {\it identical} limit cycle oscillators. We identify a parameter region where OD and a novel {\it nontrivial} AD (NT-AD) state coexist. This NT-AD state is unique in comparison with AD owing to the fact that it is created by a subcritical pitchfork bifurcation, and parameter mismatch does not support but destroy this state. We extend our study to a network of mean-field coupled oscillators to show that the transition scenario preserves and the oscillators form a two cluster state.  
\end{abstract}


\maketitle 

\section{Introduction}
\label{sec:intro}
Oscillation quenching is an emergent and intriguing phenomenon that has been the topic of extensive research in diverse fields like physics, biology, and engineering \cite{kosprep}. There are two distinct types of oscillation quenching processes:  amplitude death (AD) and oscillation death (OD). In AD coupled oscillators come to a common stable steady state which was unstable otherwise and thus form a stable homogeneous steady state (HSS) \cite{prasad1},\cite{asen,*asen1,*prasad3}. But, in the case of OD, oscillators populate different coupling dependent steady states and thus gives rise to stable inhomogeneous steady states (IHSS); in the phase space OD may coexist with limit cycle oscillations. AD is important in the case of control applications where suppression of unwanted oscillations is necessary e.g., in Laser application \cite{laser}, neuronal systems \cite{bard}, etc. On the other hand, OD is a much more complex phenomenon because it induces inhomogeneity in a rather homogeneous system of oscillators that has strong connections and importance in the field of biology (e.g., synthetic genetic oscillator \cite{kosepl,*koschaos}, cellular differentiation \cite{cell}), physics \cite{odphys}, etc.

Although, AD and OD are two structurally different phenomena--their genesis and manifestations are different, but for many years they are (erroneously) treated in the same footing. Only recently pioneering works in Ref.\cite{kosprl,kospre,kosprep} established the much needed distinctions between AD and OD (see Ref.\cite{kosprep} for an extensive review on OD). Although an extensive research work has been reported on AD (see \cite{prasad1} and references therein), but the phenomenon of OD is a less explored topic. \citet{kosprl} show that AD and OD can simultaneously occur in diffusively coupled Stuart-Landau oscillators; the authors show an important transition phenomenon, namely the transition from AD to OD in  Stuart-Landau oscillators with {\it parameter mismatch}. It established that the transition occurs due to the interplay between the heterogeneity and the coupling parameter that is analogous with the Turing-type bifurcation \cite{turing} in spatially extended systems. In \cite{kospre} it was shown that  the presence of time-delay enhances the effect of AD-OD transition; it also shows that AD-OD transition  can be induced even in the identical Stuart-Landau oscillators by using dynamic \cite{dyn}, and conjugate \cite{karna} coupling. More recently, Ref.~\cite{dana,*dana1} shows the transition between AD and OD in identical nonlinear oscillators that are coupled diffusively and perturbed by a symmetry breaking repulsive coupling link.   

In the above mentioned studies the role of mean-field diffusive coupling on the occurrence of OD, and the AD-OD transition is not considered;  mean-field coupling is one of the most widely studied topics because of its presence in many natural phenomena in the field of biology, physics, and engineering \cite{shino,*st,*de,srimali,tanchaosad,pathria}. All the previous studies show that the mean-field coupling in oscillators can induce  AD only \cite{st,de,srimali,tanchaosad}. Only in Refs.\cite{kospre2} and \cite{qstr,*qstr2,*qstr3}, in the context of genetic oscillators interacting through a quorum-sensing mechanism, the occurrence of OD is shown where the concentration of the autoinducer molecule that can diffuse through the cell membrane contains a mean-field term, but no {\it AD-OD transition} is reported there. In this paper, for the first time, we systematically explore that  the mean-field coupling can induce a Turing-type transition from AD (stable HSS) to OD (stable IHSS) even in identical limit cycle oscillators. Further, we identify an important parameter regime where OD coexists with a novel {\it non trivial} AD (NT-AD) state. This NT-AD state is unique in comparison with its conventional counterpart in, at least, two ways. Firstly, unlike AD that has two possible routes: Hopf and saddle-node bifurcation, the NT-AD state is born via a subcritical pitchfork bifurcation. Secondly, in sharp contrast with the AD, which is supported or enhanced by parameter mismatch, the NT-AD state is completely destroyed by parameter mismatch. In this paper we consider a single paradigmatic oscillator, namely Stuart-Landau oscillator, which is widely used in literature of the studies on OD, AD  and their transitions \cite{kosprl,kospre,kosprep}. We also extend our study to a network of oscillators and show that the occurrence of OD, and AD-OD transition are preserved for more than two oscillators.

\section{Stuart-Landau Oscillators with mean-field coupling}
\label{sec2}
We consider $N$ number of Stuart-Landau oscillators interacting through mean-field diffusive coupling; mathematical model of the coupled system is given by
\begin{equation}\label{ls} 
\dot{Z_i}=(1+i\omega_i-|Z_i|^{2})Z_i+\epsilon\bigg(Q\overline{Z}-Re(Z_i)\bigg),
\end{equation}
with $i=1\cdots N$; $\overline{Z}=\frac{1}{N}\sum_{i=1}^{N}Re(Z_i)$ is the mean-field of the coupled system, $Z_i=x_i+jy_i$. The individual Stuart-Landau oscillators are of unit amplitude and having eigenfrequency $\omega_i$. The coupling strength is given by $\epsilon$, and $Q$ is a control parameter that determines the density of mean-field \cite{srimali,qstr,tanchaosad} ($ 0\leqslant Q \leqslant 1$); $Q\rightarrow 0$ indicates the self-feedback case, whereas $Q\rightarrow 1$ represents the maximum mean-field density. As the limiting case we take $N=2$, and write \eqref{ls} in the Cartesian coordinate:
\begin{subequations}
\label{system}
\begin{align}
\label{x1}
\dot{x}_{1,2} &= P_{1,2}x_{1,2}-\omega_{1,2}y_{1,2}+\epsilon[Q\overline{X}-x_{1,2}],\\
\label{y1}
\dot{y}_{1,2} &= \omega_{1,2}x_{1,2}+P_{1,2}y_{1,2}.
\end{align}
\end{subequations}
Here, $P_i = 1-{x_i}^2-{y_i}^2$ $(i = 1,2)$, $\overline{X} = \frac{x_1+x_2}{2}$. At first we consider the case of two identical oscillators, i.e., $\omega_{1,2}=\omega$. From Eq.\eqref{system} it is clear that the system has the following fixed points: the trivial fixed point is the origin $(0, 0, 0, 0)$, and additionally two coupling dependent nontrivial fixed points: (i) (${x_1}^\ast$, ${y_1}^\ast$, $-{x_1}^\ast$, $-{y_1}^\ast$) where ${x_1}^\ast = -\frac{\omega {y_1}^\ast}{{\omega}^2 + \epsilon {{y_1}^\ast}^2}$ and ${y_1}^\ast = \sqrt {\frac{(\epsilon - 2{\omega}^2) + \sqrt{{\epsilon}^2 - 4{\omega}^2}}{2\epsilon}}$. (ii) (${x_1}^\dagger$, ${y_1}^\dagger$, ${x_1}^\dagger$, ${y_1}^\dagger$) where ${x_1}^\dagger = - \frac{\omega {y_1}^\dagger}{\epsilon(1 - Q){{y_1}^\dagger}^2 + {\omega}^2}$ and ${y_1}^\dagger = \sqrt{\frac{\epsilon(1 - Q) - 2 {\omega}^2 + \sqrt{{(\epsilon - \epsilon Q)}^2 - 4{\omega}^2}}{2\epsilon(1 - Q)}}$.

Note that, the existence of these nontrivial fixed points was not explored in the earlier study of mean-field coupled Stuart-Landau oscillators \cite{srimali}. In the next sections we will examine different dynamical regions and their transitions based on the eigenvalue analysis; subsequently, we carry out bifurcation analysis using the package $\mbox{XPPAUT}$ \cite{xpp}. 

\section{AD-OD transition and emergence of nontrivial AD}
\label{sec3}
The four eigenvalues of the system at the trivial fixed point $(0,0,0,0)$ are,
\begin{subequations}
\label{lambda}
\begin{align}
\label{lambda1}
{\lambda}_{1,2} &= 1 - \left[\frac{\epsilon(1-Q)\pm\sqrt{{\epsilon}^2(1-Q)^2-4{\omega}^2}}{2}\right],\\
\label{lambda3}
{\lambda}_{3,4} &= 1- \left[\frac{\epsilon \pm \sqrt{{\epsilon}^2-4{\omega}^2}}{2}\right].
\end{align}
\end{subequations}
Eigenvalue analysis and also a close inspection of the nontrivial fixed points reveal that the system has two pitchfork bifurcations (PB) given by PB1 and PB2 occurring at the following values of the coupling parameters, respectively:
\begin{subequations}
\label{pb}
\begin{align}
\label{epsapb1}
{\epsilon}_{PB1} &= 1+{\omega}^2,\\
\label{epsapb2}
{\epsilon}_{PB2} &= \frac{1+{\omega}^2}{1-Q}.
\end{align}
\end{subequations}
${\epsilon}_{PB1}$ is that value where a symmetry breaking pitchfork bifurcation gives birth to the nontrivial fixed point (${x_1}^\ast$, ${y_1}^\ast$, $-{x_1}^\ast$, $-{y_1}^\ast$), i.e., IHSS emerges at this value of coupling parameter. It is noteworthy that the occurrence of PB1 does not depend upon the density parameter $Q$ (but, later we will see that stability of IHSS depends on $Q$). The second nontrivial fixed point (${x_1}^\dagger$, ${y_1}^\dagger$, ${x_1}^\dagger$, ${y_1}^\dagger$) arises at PB2; PB2 gives rise to an unique {\it nontrivial HSS} state. Later we will see that stabilization of this state leads to a novel {\it nontrivial} AD (NT-AD) state that coexists with OD.  

Next, we search for the Hopf bifurcation point at which the stable oscillation dies to give birth to AD state. From \eqref{lambda} it is clear that for $\omega\le1$ no Hopf bifurcations (of trivial fixed point) occur, only pitchfork bifurcations govern the dynamics in that case. For any $\omega>1$, equating the real part of ${\lambda}_{3,4}$ and ${\lambda}_{1,2}$ to zero we get, 
\begin{subequations}
\label{epsahb}
\begin{align}
{\epsilon}_{HB1} &= 2,\\
{\epsilon}_{HB2} &= \frac{2}{1-Q},
\end{align}
\end{subequations}
respectively; here ${\epsilon}_{HB1}$ and ${\epsilon}_{HB2}$ are the values of coupling parameters where first (HB1) and second (HB2) Hopf Bifurcation occur, respectively. From (\ref{epsahb}) it is clear that ${\epsilon}_{HB1}$ is constant, but ${\epsilon}_{HB2}$ depends only upon $Q$ value (and independent of $\omega$, where $\omega>1$). Now, when $Q \rightarrow 0$, ${\epsilon}_{HB1} \approx {\epsilon}_{HB2}$.
\begin{figure}
\includegraphics[width=.45\textwidth]{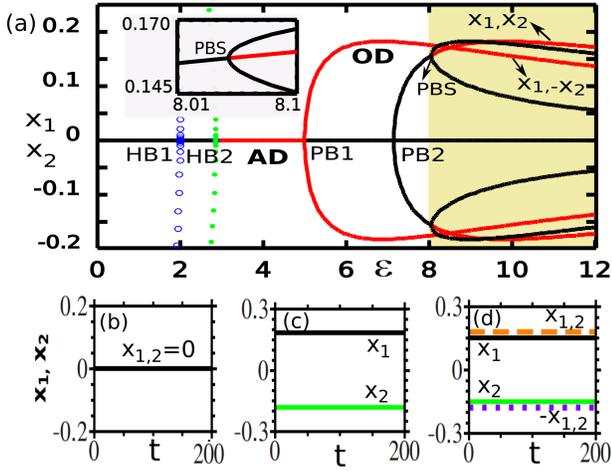}
\caption{\label{fig1}(Color online) (a) Bifurcation diagram (using $\mbox{XPPAUT}$) of two mean-field coupled identical Stuart-Landau oscillators ($Q=0.3$, $\omega=2$). Grey (red) lines: stable fixed points, Black lines: unstable fixed points, solid circle (green): stable limit cycle, open circle (blue): unstable limit cycle. HB1,2 and PB1,2 are Hopf and pitchfork bifurcation points, respectively. PBS denotes subcritical pitchfork bifurcation point; inset shows the zoomed in view of the region of occurrence of PBS. AD is created at HB2, and PB1 gives the AD-OD transition point. Coexistence of OD ($x_1=-x_2$) and {\it nontrivial} AD (NT-AD) ($x_1=x_2$) is shown in shaded (yellow) region. Time traces are shown for (b) AD ($x_{1,2}=0$) at $\epsilon=4$, (c) OD ($x_1=-x_2$) at $\epsilon=7$, and (d) NT-AD and OD at $\epsilon=10.92$; here dashed and dotted lines represent two initial condition dependent NT-AD states, $x_{1,2}$ and $-x_{1,2}$, respectively.}
\end{figure}
Figure~\ref{fig1} (a) shows the bifurcation diagram of $x_{1,2}$ for $Q=0.3$ and $\omega=2$ (without any loss of generality, unless stated otherwise, we take $\omega=2$). It is observed that at HB2 an inverse Hopf bifurcation occurs and the stable limit cycle is suppressed to give birth of AD (i.e., a stable HSS state); whether at HB1 an unstable limit cycle is born. This stable HSS (AD) state becomes unstable trough a supercritical pitchfork bifurcation (PB1) at ${\epsilon}_{{PB}_1}= 1+{\omega}^2=5$. Here the trivial fixed point becomes unstable and two new stable IHSSs are created giving birth to OD. Thus, we get a transition between AD and OD in identical mean-field coupled oscillators. With further increase in coupling strength ($\epsilon$), PB2 occurs at $\epsilon_{PB2}=7.142$ (that agrees with \eqref{epsapb2}), which gives birth to a {\it nontrivial} HSS (i.e., ${x_1}^\dagger={x_2}^\dagger$). This {\it nontrivial} HSS is stabilized via subcritical pitchfork bifurcation at $\epsilon_{PBS}\approx8.05$ and gives rise to a novel {\it nontrivial} AD (NT-AD) state. We attach the attribute {\it nontrivial} to this AD state because it emerges from the {\it nontrivial} HSSs ($x^\dagger,y^\dagger$), which are non-zero and subsequently placed symmetrically around zero.  We also verify the occurrence of this pitchfork bifurcation directly from the eigenvalues corresponding to $({x_1}^\dagger$, ${y_1}^\dagger$, ${x_1}^\dagger$, ${y_1}^\dagger)$, which are given by: 
\begin{subequations}
\label{ntlambda2}
\begin{align}
\label{ntlambda12}
{\lambda}^{\dagger}_{1,2} &= 1 - \frac{{b_1}^\dagger}{2} \pm \frac{\sqrt{{{b_1}^\dagger}^2-4{c_1}^\dagger}}{2}, \\
\label{ntlambda34}
{\lambda}^{\dagger}_{3,4} &= 1 - \frac{{b_2}^\dagger}{2} \pm \frac{\sqrt{{{b_2}^\dagger}^2-4{c_2}^\dagger}}{2}, 
\end{align}
\end{subequations}  
where ${b_1}^\dagger = (\epsilon -\epsilon Q+4{{x_1}^\dagger}^2+4{{y_1}^\dagger}^2)$, ${c_1}^\dagger =({{x_1}^\dagger}^2+3{{y_1}^\dagger}^2)(\epsilon -\epsilon Q+3{{x_1}^\dagger}^2+{{y_1}^\dagger}^2)+{\omega}^2-4{{x_1}^\dagger}^2{{y_1}^\dagger}^2$, ${b_2}^\dagger = (\epsilon+4{{x_1}^\dagger}^2+4{{y_1}^\dagger}^2)$, ${c_2}^\dagger = ({{x_1}^\dagger}^2+3{{y_1}^\dagger}^2)(\epsilon+3{{x_1}^\dagger}^2+{{y_1}^\dagger}^2)+{\omega}^2-4{{x_1}^\dagger}^2{{y_1}^\dagger}^2$. Since stable IHSS (OD) solutions [corresponding to the first nontrivial fixed points ($x^\ast, y^\ast$)] still exist beyond this coupling value thus OD and NT-AD coexist for $\epsilon\ge\epsilon_{PBS}$ [shaded (yellow) region in Fig.~\ref{fig1} (a)]. Coexistence of OD and another kind of nontrivial AD was found earlier in conjugate coupled Stuart-Landau oscillators in \cite{kospre}, but here the genesis of NT-AD and the origin of coexistence is different from that; in our case subcritical pitchfork bifurcation is responsible for the NT-AD state. Further, in the NT-AD state we have two different solutions: $x_1=x_2$, and $-x_1=-x_2$; the occurrence of one of these two states is determined by the initial conditions. This has a striking resemblance to bi-stability, but here the bi-stability is much more subtle owing to the fact that, unlike its classical counterpart, it coexists with OD and it emerges via a subcritical pitchfork bifurcation. Later we will see that any parameter mismatch destroys this NT-AD state. This initial condition dependent amplitude death state is not observed earlier. To confirm the coexistence of OD and NT-AD we integrate the system equation  with suitably chosen initial conditions (using fourth-order Runge-Kutta method; step size $=0.005$); Fig.\ref{fig1}(d) shows this for $\epsilon=10.92$, where we can see the OD state and NT-AD states coexist. Figure \ref{fig1}(b) and (c) show the AD ($\epsilon=4$) and OD ($\epsilon=7$) states, respectively.

\begin{figure}
\includegraphics[width=.45\textwidth]{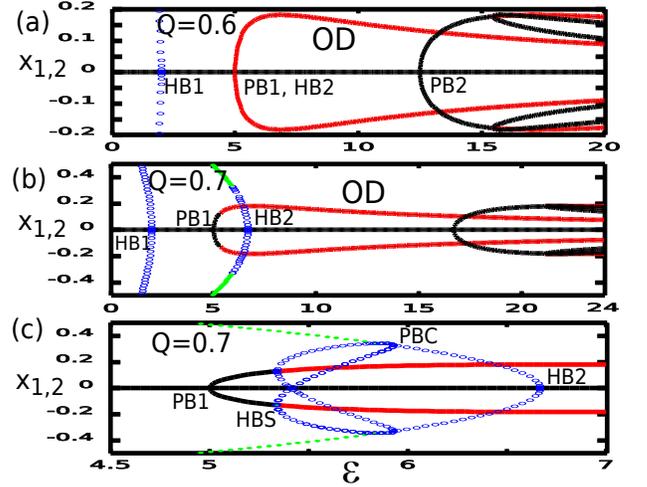}
\caption{\label{fig2} (Color online) (a) AD state vanishes as $HB2=PB1$ at $Q=Q^\ast$ (=0.6) (b), (c) $Q>Q^\ast$ (=0.7): HB2 moves to the right side of PB1, nontrivial fixed point gets stability by subcritical Hopf bifurcation (HBS). Between HBS and PBC (pitchfork bifurcation of limit cycle) coexistence of stable, unstable limit cycle, and OD is observed. Other parameter: $\omega=2$.}
\end{figure}

Now, with increasing $Q$ value, ${\epsilon}_{HB2}$ will move towards ${\epsilon}_{PB1}$ and the zone of stable HSS (AD) reduces. For a given $\omega$ (where $\omega>1$) at a particular $Q$ value (say $Q^\ast$), ${\epsilon}_{HB2}$ will collide with ${\epsilon}_{PB1}$. So at $Q=Q^\ast$, ${\epsilon}_{HB2} = {\epsilon}_{PB1}$, i.e., $ Q^\ast = \frac{{\omega}^2-1}{{\omega}^2+1}$. At this point, the $\epsilon$ region where AD occurs vanishes and thus AD to OD transition does not occur. Figure~\ref{fig2}~(a) shows this scenario for $\omega=2$, and $Q=0.6$. Now, for $Q>Q^\ast$, ${\epsilon}_{HB2} > {\epsilon}_{PB1}$, i.e., HB2 point moves towards the right hand side of PB1; subsequently, the IHSS now gains stability at $\epsilon_{HBS}$ through a subcritical Hopf bifurcation; in Fig.\ref{fig2} (b) and (c) for $Q=0.7$ we get $\epsilon_{HBS}\approx5.341$ . This can be predicted from the eigenvalues of the nontrivial fixed point (${x_1}^\ast$, ${y_1}^\ast$, $-{x_1}^\ast$, $-{y_1}^\ast$), which are same as \eqref{ntlambda2} but with the ($\dagger$) signs replaced by ($\ast$) signs.
\begin{figure}
\includegraphics[width=.35\textwidth]{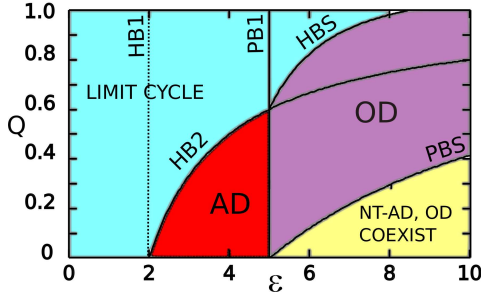}
\caption{\label{twop} (Color online) Phase diagram in $Q-\epsilon$ space ($\omega=2$). With increasing $Q$, collision of HB2 and PB1 destroys the AD-OD transition scenario.}
\end{figure}
From the eigenvalue equations we find  ${\epsilon}_{{HBS}}$ where the IHSS regains stability: 
\begin{equation}
\label{epsahb3}
{\epsilon}_{HBS} = \frac{-2(Q+1)+4\sqrt{1+{\omega}^2(1-Q)(3+Q)}}{(1-Q)(3+Q)}.
\end{equation}
The value of ${\epsilon}_{HBS}$ agrees with Fig.\ref{fig2} (b) and (c). HB2 point gives birth to an unstable limit cycle that becomes stable through a pitchfork bifurcation of limit cycle (PBC). Between HBS and PBC, stable and unstable limit cycles coexist with OD. In this region we identify (not shown here) three distinct dynamical behaviors: homogeneous limit cycle (HLC), inhomogeneous limit cycle (IHLC), and OD. We grab the whole bifurcation scenario in the  $Q-\epsilon$ parameter space (Fig.~\ref{twop}). We can see that, with increasing $Q$, at $Q=0.6$, HB2 collides with PB1, thus destroying the AD-OD transition. It also shows the coexisting region of NT-AD and OD that is determined by the PBS curve. In the previous studies on the mean-field coupled Stuart-Landau oscillators only the transition from limit cycle to AD was shown \cite{srimali}; here we identify additional bifurcation scenarios and dynamical regions. Before we proceed further let us summarize our results of AD-OD transition: (i) For $Q<Q^\ast$, ${\epsilon}_{HB2} < {\epsilon}_{PB1}$: AD-OD transition occurs. (ii) For $Q=Q^\ast$, ${\epsilon}_{HB2} = {\epsilon}_{PB1}$: No AD, only stable IHSS (OD); AD-OD transition vanishes. (iii) For $Q>Q^\ast$, ${\epsilon}_{HB2} > {\epsilon}_{PB1}$: IHSS gains stability at $\epsilon_{HBS}$, OD moves to the right hand side with increasing $Q$.
\section{Parameter mismatch, Cluster formation}
\label{sec4}
\begin{figure}
\includegraphics[width=.35\textwidth]{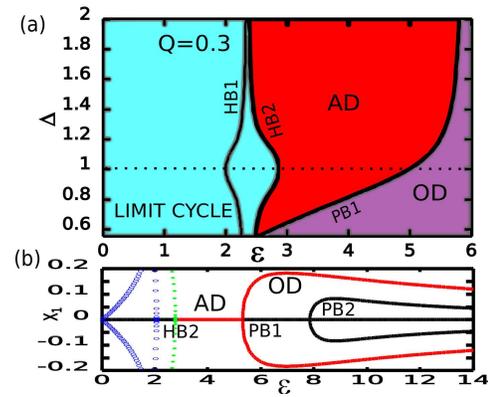}
\caption{\label{2pdelta} (Color  online) (a) Phase diagram in $\Delta-\epsilon$ space for $Q=0.3, \omega_1=2$. NT-AD state vanishes for any $\Delta\ne1$; this is shown in (b) for $\Delta=1.1$.}
\end{figure}
\begin{figure}
\includegraphics[width=.45\textwidth]{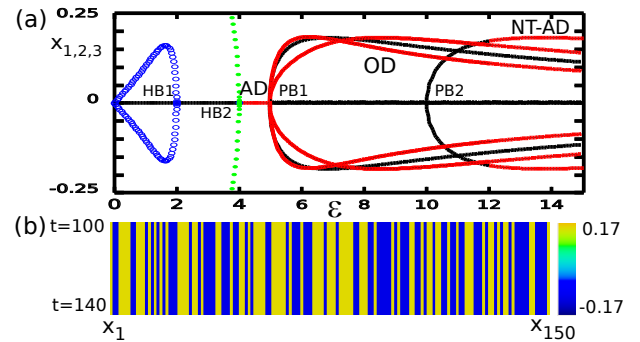}
\caption{\label{st} (Color online) (a) Bifurcation for $N=3$ ($\omega=2$): AD-OD transition is preserved; also, NT-AD state ($\alpha,\alpha,\alpha$) coexists with OD. (b) Two-cluster pattern formation: Space-time plot of network of 256 (150 are shown for clarity) mean-field coupled Stuart-Landau oscillators at $\epsilon=16$, $\omega=3$. Other parameters: $Q=0.5, \Delta=1$.}
\end{figure}
We examine the effect of parameter mismatch on the coupled dynamics. We introduce a mismatch parameter $\Delta$ in Eq.\eqref{system} defined by $\Delta=\omega_2/\omega_1$. $\Delta=1$ represents the case of no mismatch. For $\Delta\ne1$, nontrivial fixed points of \eqref{system} can not be derived in a closed form, thus we use $\mbox{XPPAUT}$ to locate them, and subsequently test their stability. To get a detail scenario of the dynamical behaviors we compute the two-parameter bifurcation diagram in $\Delta-\epsilon$ space for a given $Q$ and $\omega_1$. Figure \ref{2pdelta} (a) shows this for $Q=0.3$ and $\omega_1=2$. It can be observed that for the mismatched case AD occurs at lower value of $\epsilon$. It is noteworthy that the HB2 curve is symmetrical around $\Delta=1$ line; this is expected as HB2 does not depend upon $\omega$ (as long as $\omega>1$). OD is governed by the PB1 curve, which depends upon the frequency of oscillators and thus on the value of $\Delta$. For $\Delta<1$, PB1 comes closer to HB2 and thus reducing the zone of AD, and broadening the zone of OD. At $\Delta\approx0.54$, PB1 and HB2 collide to eliminate the zone of AD, and thus destroy the AD-OD transition. For $\Delta>1$, PB1 moves far from HB2 enhancing the zone of AD, and also supporting the AD-OD transition. Thus, we see that beside $Q$, AD-OD transition is determined by the parameter mismatch, also. We have made another important observation in the mismatched case: the {\it nontrivial} HSS created at PB2 does not get stable for any $\Delta\ne1$. Thus, for the parameter mismatched case no NT-AD state occurs.  As an illustrative example, Fig. \ref{2pdelta} (b) shows that no NT-AD occurs making OD the only possible solution beyond PB1 ($\Delta=1.1$, $Q=0.3$, $\omega_1=2$). Nevertheless, the {\it nontrivial} HSS (although unstable) still exists even in the parameter mismatched case.

Next, we investigate the more general case of $N>2$. At first let us take $N=3$ and $\Delta=1$; now the coupled equation is given by Eq.\eqref{ls} with $i=1,2,3$. Beside the trivial fixed point, there exist other nontrivial solutions with the combination like, ($\alpha,\alpha,\alpha$), ($\alpha,\beta,\alpha$), ($\alpha,\alpha,\beta$) and their cyclic permutations \cite{kospre}. The ($\alpha,\alpha,\alpha$) set gives the {\it nontrivial} HSS and the remaining sets give IHSS solutions. Fig.\ref{st} (a) shows this scenario for $\omega=2$ and $Q=0.5$. Here also, we can observe the occurrence of AD-OD transition, and coexistence of OD and NT-AD. Next, we consider the network of $N=256$ mean-field coupled oscillators; Fig.\ref{st} (b) shows the space-time plot of stable IHSS (OD) solutions for $\Delta=1$, $\epsilon=16$, $\omega=3$ and $Q=0.5$ (for clarity the first 150 elements are shown). The figure clearly shows the formation of a two-cluster solution. Further, we observed that (not shown here) in the space-time plot the size and position of the domains change with the number of elements ($N$) and initial conditions; clearly this fact has a striking resemblance with the {\it frozen random pattern} solution of a coupled map lattice system \cite{kan2,*tanchaos}. 
\section{\label{con}Conclusion}
\label{sec5}
We have explored the phenomena of AD, OD and their transitions in the paradigmatic Stuart-Landau oscillators under the mean-field diffusive coupling. Using detailed eigenvalue analyses supported by bifurcation analyses we have shown that the mean-field diffusive coupling can induce OD and also a transition between AD and OD even in identical Stuart-Landau oscillators. It has been shown that while the presence of mean-field density parameter is not essential for inducing OD, but the AD-OD transition is absolutely governed by the mean-field density parameter; the relevance of this parameter was discussed earlier in the context of genetic oscillators interacting through a quorum-sensing mechanism \cite{qstr}. We have identified a novel dynamical state that is created by subcritical pitchfork bifurcation, namely {\it nontrivial} AD (NT-AD) that coexists with the OD region. Unlike (conventional) AD this state is destroyed by the presence of parameter mismatch. Further, in the NT-AD state the occurrence of one of the two states is determined by the initial conditions; to the best of our knowledge, this initial condition dependent amplitude death state has not been observed earlier. However, the observation of NT-AD is subtle in natural and experimental systems as parameter mismatch is inevitable in the practical coupled oscillators \cite{tanexptod}. We have also extended our findings to a network of identical mean-field coupled Stuart-Landau oscillators where it has been shown that the AD to OD transition scenario is preserved; in this case we have shown that the coupled oscillators form a two-cluster state, population of which depends upon the initial conditions.  This study can be extended to other limit cycle and chaotic oscillators and we believe that this will improve our understanding of various mean-field coupled biological and engineering systems.

\providecommand{\noopsort}[1]{}\providecommand{\singleletter}[1]{#1}%
\end{document}